\documentclass[11pt]{article}
\usepackage{moriond,epsfig}

\def\Journal#1#2#3#4{{#1} {\bf #2}, #3 (#4)}

\def\ZPC{{\em Z. Phys.} C}
\def\JGR{\em J. Geophys. Res.}
\def\AP{\em Astropart. Phys.}

\def\be{\begin{equation}}
\def\ee{\end{equation}}
\def\bea{\begin{eqnarray}}
\def\eea{\end{eqnarray}}

\begin{document}
\vspace*{4cm}
\title{NEUTRINO ASTRONOMY AT THE SOUTH POLE}

\author{P.A. TOALE \footnote{On behalf of the IceCube Collaboration, see
        {\sf http://icecube.wisc.edu/science/publications/moriond2006.html}.
	\label{foot}} }

\address{The Pennsylvania State University\\
         Department of Physics\\ 
         104 Davey Lab, PMB046\\
         University Park, PA, 16802}

\maketitle\abstracts{
IceCube is currently being built deep in the glacial ice beneath the South Pole. 
In its second year of construction, it is already larger than its predecessor,
AMANDA. 
AMANDA continues to collect high energy neutrino and muon data as an independent
detector until it is integrated with IceCube.
After introducing both detectors,
recent results from AMANDA and a status report on IceCube are presented.}

\section{Introduction}
Traditional astronomical observations utilize photons or charged particles.
Photons are detected over a wide range of energies, from the 2.7 K cosmic
microwave background up to $10~\mbox{TeV}$ gamma rays. 
Unfortunately photon astronomy is not possible at
higher energies due to interactions with interstellar gases and radiation
fields. Cosmic rays are 
detected at even higher energies, and the energy spectrum has been  
measured out as high as $10^{11}$ GeV. However, all but the highest energy rays
(above $10^{10}$ GeV) are bent in the galactic magnetic field, obscuring the
identity of their sources.

There are a variety of models that account for the high energy photons and 
charged particles that are observed. 
These can be divided into two categories: hadronic acceleration and
exotic particle decay. The first category contains the so--called bottom--up
models since they all involve acceleration of hadrons, typically 
protons, to very
high energy. Candidates for these cosmic accelerators include Active Galactic
Nuclei, Supernova remnants, and Gamma Ray Bursts. Models of the second category
are known as top--down since they contain a heavy exotic particle 
decaying into energetic 
Standard Model particles. Examples include topological defects such
as magnetic monopoles and generic Weakly Interacting Massive Particles (WIMPs).

Most of these models predict neutrino production in addition to
photons and cosmic rays. The goal of neutrino astronomy is to discover and
measure these high energy neutrinos. Neutrinos offer several advantages over
traditional astronomical messengers. First, they are weakly interacting, so they
can travel cosmological distances without being scattered or absorbed. Second,
they are electrically neutral, so they are not deflected by interstellar
magnetic fields. Together, these two qualities make neutrinos excellent
discovery instruments.

The same properties that make neutrinos good astronomical messengers also make
them difficult to detect. The most successful technique for observing
extraterrestrial neutrinos works by detecting Cherenkov light produced by
secondary particles arising from neutrino interactions. The signature of 
light differs
depending on the flavor of the neutrino and the type of interaction. For 
instance, muon neutrinos can interact via a charged current (CC) interaction
and produce a long lived muon with a conical light pattern, 
whereas CC interactions of electron neutrinos produce electromagnetic showers
with a more spherical pattern. Tau neutrinos produce tau leptons in CC
interactions which, depending on the size of the detector, may decay back into
tau neutrinos plus other particles, resulting in a combination of track--like
and shower--like patterns. The neutral current (NC) interaction of all
three flavors result in hadronic showers, which are probably not
distinguishable from electromagnetic showers.

In order to detect the fluxes predicted by accepted cosmological models, one
needs either to wait a very long time or to build a very big detector. This
leads to the search for naturally occurring, optically transparent media.
Several groups have implemented the optical Cherenkov technique in large 
volumes of water or ice. The two media differ both in optical properties and in
inherent backgrounds. Water typically has a long scattering length 
($> 100~\mbox{m}$) but a short absorption length ($\sim 20~\mbox{m}$). Deep 
ice~\cite{ice} on
the other hand, has a long absorption length ($\sim 110~\mbox{m}$) but a
short scattering length ($\sim 20~\mbox{m}$). Also, ice is nearly free of 
light--producing backgrounds while water contains biological and chemical 
sources such as potassium decay in salt water. The rest of this paper focuses
on neutrino detection in deep ice
as pioneered by the Antarctic Muon and Neutrino Detector Array (AMANDA) 
and carried on today by IceCube.

\section{AMANDA}

AMANDA has applied the optical Cherenkov technique
using deep, clear ice below the Amundsen--Scott South Pole station. The current
detector consists of 19 instrumented strings, deployed between
$1500-2000~\mbox{m}$ below the ice surface. Each string contains between 20 to 42 
optical modules (OMs), spaced vertically by $10-20~\mbox{m}$. There are a total 
of 677 OMs in the ice, each of which is composed of an 8--inch
photomultiplier tube (PMT) housed in a glass pressure sphere. The PMT signals
are sent to the surface over either twisted pair, coaxial, or fiber
optic cable where they are processed by the surface data acquisition system. The
instrumented region of AMANDA is cylindrical in shape with a $200~\mbox{m}$ 
diameter and a height of $500~\mbox{m}$.

The AMANDA detector is best suited for detection of muons. These can be 
atmospheric muons produced from cosmic rays or muons produced from muon 
neutrino interactions. These muon neutrinos can arise from cosmic
rays interaction with the atmosphere or from the much anticipated
cosmological progenitors. High energy muons
travel long distances through the ice, {\it e.g.} a $200~\mbox{GeV}$ muon 
can travel up to $1~\mbox{km}$. 
At the depth of AMANDA (and IceCube), the rate of down-going
atmospheric muons is a million times higher than the rate of muons from
atmospheric neutrinos. Therefore the characteristic signature of muon neutrinos
in AMANDA is muon tracks traveling upwards through the detector toward the
surface.

The direction (space--angle) of muon tracks passing through the detector can
be reconstructed with an accuracy of $2-3^\circ$. Since the direction of the
primary muon neutrino differs from the direction of the muon by
$\sim 1^\circ$ at $1~\mbox{TeV}$, the neutrino points back toward its source with
similar accuracy. The energy reconstruction is much less precise because it is
based on radiative losses of the muon as it passes through the detector.
Fluctuations in these losses result in an energy resolution of
30\% in the logarithm of energy.

Cascade reconstruction is significantly different. Light from cascades is
produced over $\sim 10~\mbox{m}$ which, given the spacing of the OMs, results in
a degraded angular resolution of $30-40^\circ$. On the other hand, cascades are
likely to be contained within the detector, giving them a better energy
resolution (as good as 10\% in $\log{E}$).

\subsection{A Selection of AMANDA Results}
AMANDA analyses fall into three broad categories: measurement of the energy
spectrum of the diffuse neutrino flux, searches for neutrino point sources, 
and search for new physics in the form of exotic particles or violation of
conservation laws. A summary of each of the categories is given below.

The diffuse flux is studied by first measuring the energy spectrum of 
up-going muons and then unfolding resolution effects to arrive at the neutrino
energy spectrum. The spectrum has been measured from $1-100~\mbox{TeV}$ 
and agrees well with both atmospheric models and lower energy data from the
Frejus experiment~\cite{kd}. Preliminary results from the AMANDA data 
give a spectral index of $3.54 \pm 0.11$~\cite{spec}.
There is no indication of excess events above the expected
atmospheric contribution in this energy range.

\begin{figure}
\begin{center}
\epsfig{figure=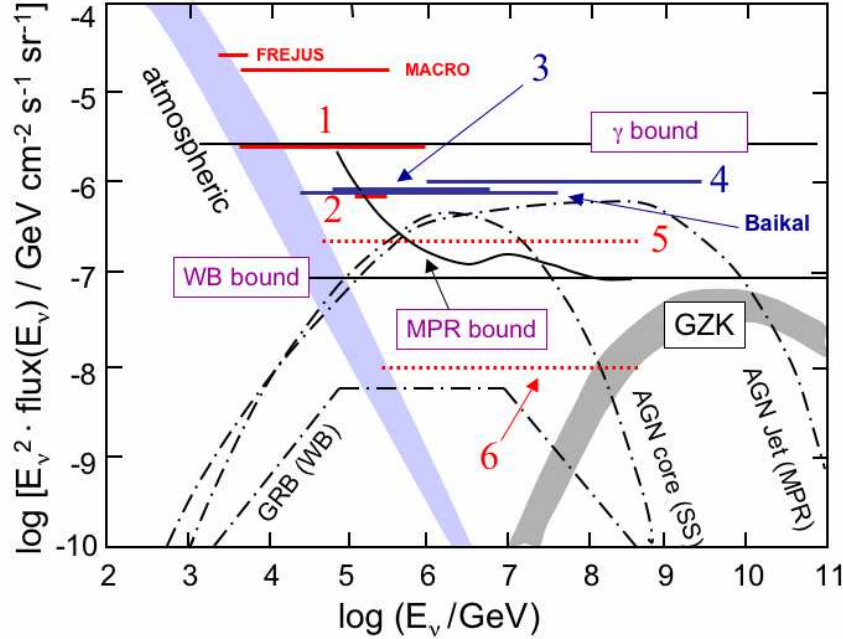,height=3.5in}
\caption{Diffuse neutrino flux limits shown with expected atmospheric 
contribution and several leading cosmological models. The limits are calculated
assuming oscillations with full mixing, so the $\nu_\mu$ limits are shown 
multiplied by 3.
\label{fig:diffuse}}
\end{center}
\end{figure}

Figure~\ref{fig:diffuse} shows upper limits and sensitivities from several 
independent analyses, along with predictions from several leading 
models. Limits (1), (3), and (4) are published results from three independent 
AMANDA analyses~\cite{l1,l3,l2}. Limit (2), along with sensitivities (5) and
(6), are preliminary AMANDA results~\cite{l4,l5,l6}. Also shown are limits from 
Frejus~\cite{kd}, MACRO~\cite{macro}, and Baikal~\cite{baikal}.

\begin{figure}
\begin{center}
\epsfig{figure=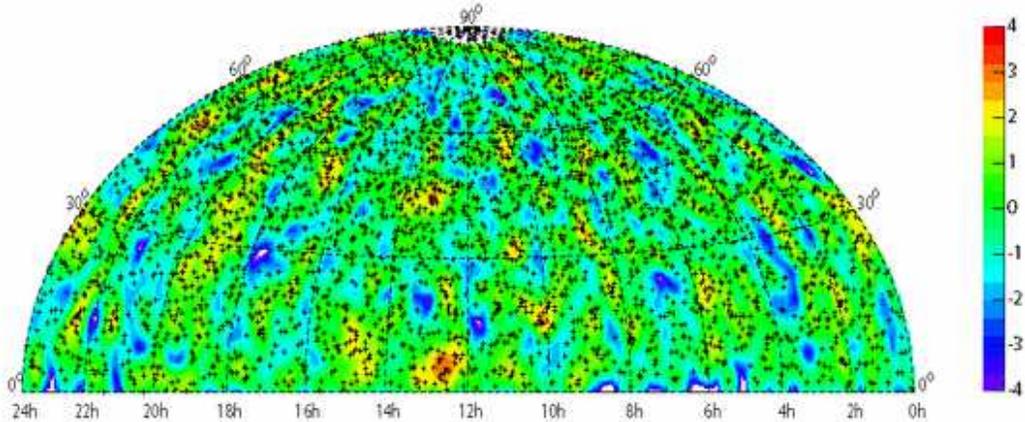,width=6.0in}
\caption{Northern hemisphere of the neutrino sky map based on 3329 muon neutrino
events. Colors indicate Gaussian standard deviations from the expected isotropic
distribution.
\label{fig:ps}}
\end{center}
\end{figure}

The point--source analysis uses 3329 up--going events collected from 2000 to
2003~\cite{icrc}. The resulting, preliminary, sky map, shown in 
Figure~\ref{fig:ps},
is consistent with isotropic atmospheric neutrinos.
The maximum observed deviation is $3.4 \sigma$, which has a chance occurrence of
92\%. A catalog of 33 known astronomical sources is also studied, of which the
most significant excess is observed from the Crab nebula with 10 events and
an expected background of 5.4. The chance probability of this excess is 64\%. 

AMANDA continues to search for dark matter in the form of neutralinos. In many
models the neutralino is the lightest supersymmetric particle and is stable if
R--parity is conserved. Over time, these particles will accumulate at the
bottom of gravitational wells, such as the center of the sun. The rate at which
they collect depends on several factors, such as their density and velocity
distributions, their cross sections for interacting with ordinary matter, and
their annihilation cross section. The annihilation products include $q\bar{q}$, 
$l\bar{l}$, as well as $W^\pm$, $Z$, and $H$, with $b\bar{b}$, $\tau^+\tau^-$,
and $W^+W^-$ dominating. Many of the secondary decays include neutrinos.

\begin{figure}
\begin{center}
\epsfig{figure=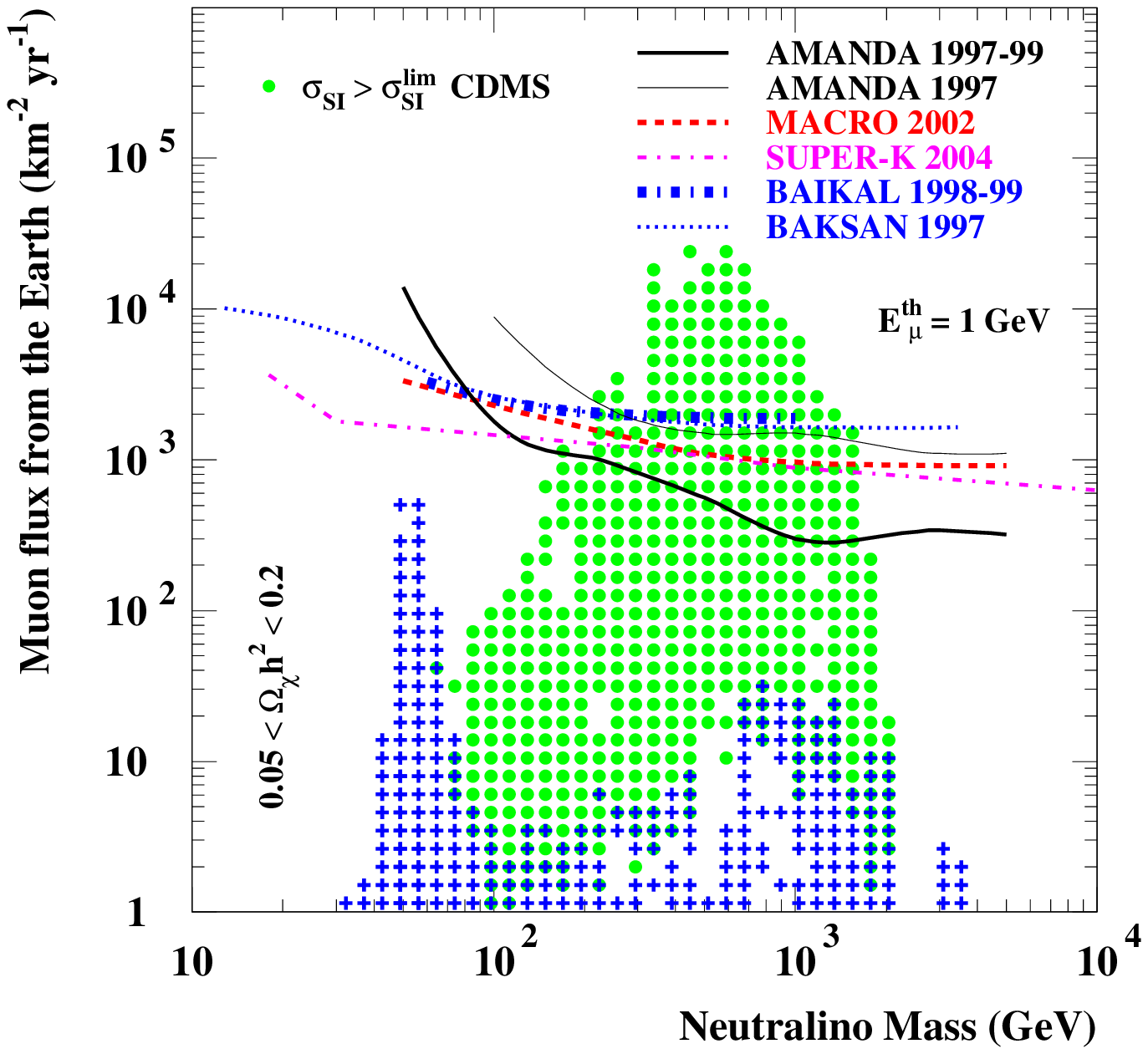,height=3.0in}
\epsfig{figure=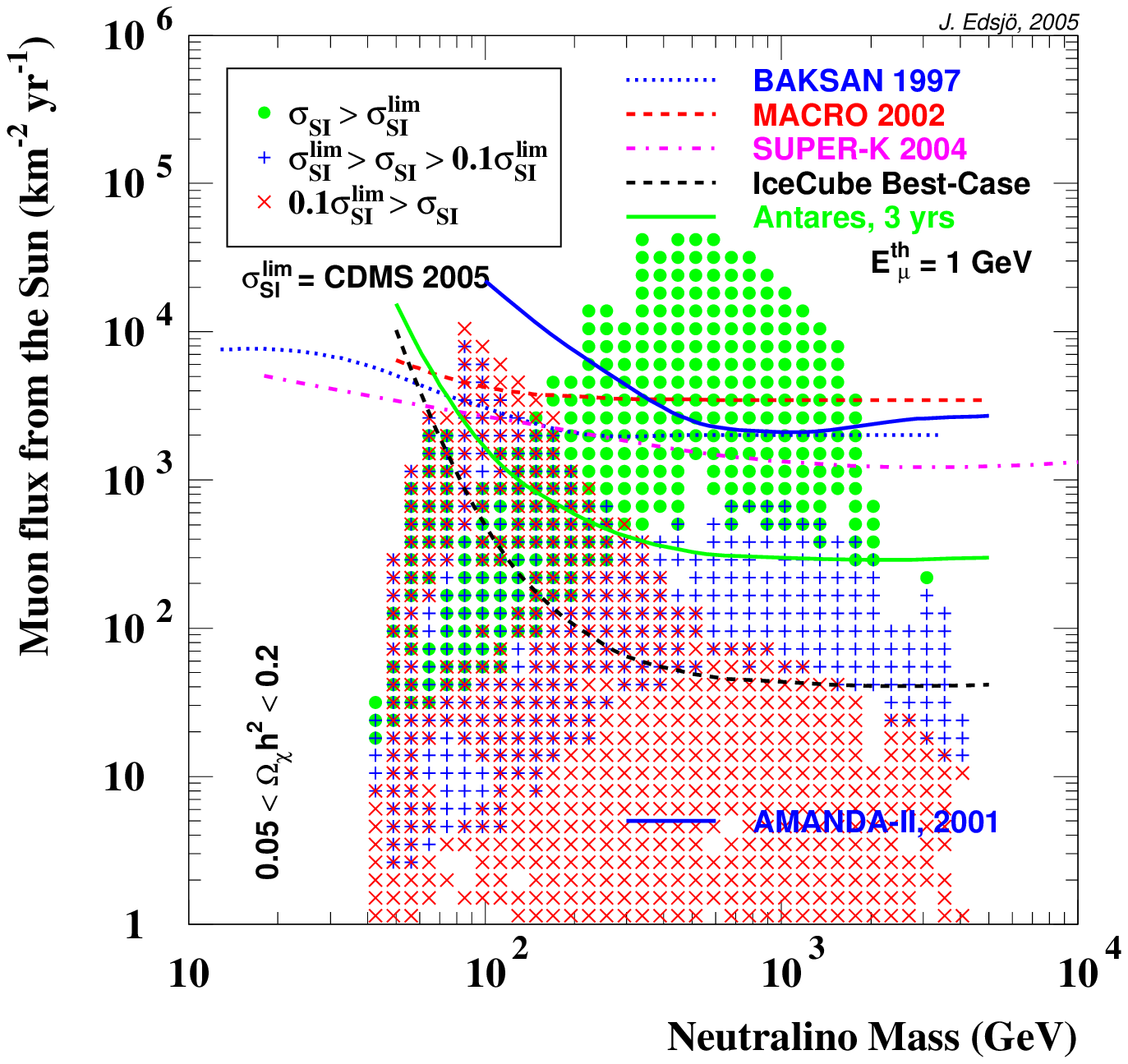,height=3.0in}
\caption{Upper limits on mass dependent muon neutrino fluxes from the center of
the earth (left) and the center of the sun (right). In both cases the CDMS
results are shown as the hashed region with the green area laying within
$1~\sigma$ of their spin--independent limit. 
\label{fig:wimps}}
\end{center}
\end{figure}

The analysis is therefore optimized to detect neutrinos produced in the sun or
in the center of the earth. So far, only the muon neutrino channel has been
probed. No excess has been observed from either source. 
Figure~\ref{fig:wimps} shows exclusion plots of muon flux as a function of
neutralino mass. In both, the hashed region shows the sensitivity of the CDMS
experiment, with the green area excluded at the one sigma level.
The plot on the left shows the upper limit from the center-of-earth
analysis~\cite{earth}, while the one on the right shows the results from the
center-of-sun analysis~\cite{solar}.

\section{IceCube}
IceCube differs from AMANDA in two important ways. First, IceCube will
be much larger, with a final fiducial volume of approximately one cubic
kilometer. Second, PMT signals are
digitized in situ by electronics housed in the optical module. 

IceCube will eventually consist of 70 or more strings with 60 modules each. The
modules are spaced by $17~\mbox{m}$ vertically while the strings are arranged in
a hexagonal pattern with a spacing of $125~\mbox{m}$. In addition to the in-ice
array, there is a surface cosmic ray detector named IceTop. At the position of
each in-ice string there is a surface station comprised of two tanks, 
each containing two DOMs frozen in clear ice.

The heart of IceCube is the Digital Optical Module (DOM) that consists of a
$10~\mbox{in.}$ PMT together with fast digitization electronics. Signals are
processed by both an Advanced Transient Waveform Digitizer (ATWD)
and a long running Fast Analog-to-Digital Converter (FADC). The ATWD contains 
three channels with different gains and 
has a programmable sampling rate from $200-700~\mbox{MHz}$ for a
maximum range of $400~\mbox{ns}$. The FADC, on the other hand, samples at a
fixed rate of $40~\mbox{MHz}$ but has a range of $6.4~\mu\mbox{s}$. 

The dark noise rates of the DOM in ice is $650~\mbox{Hz}$. This rate is
reduced in the ice by local coincidence logic which makes use of information
from neighboring DOMs. By requiring activity in a neighbor in a $800~\mbox{ns}$
time window, the rate is reduced to $15~\mbox{Hz}$ on average.

Construction of IceCube began in the Austral summer of 2004-2005 with the
deployment of a single in--ice string and four surface stations and continued 
in 2005-2006 with eight strings and 12 stations. 
Construction is scheduled to continue through 2010.

\subsection{IceCube First Year Performance}
IceCube collected data with its first string (String 21) of modules throughout
2005. Studies of initial performance~\cite{fy} indicate that the design
goals have been met. 

In particular, the average dark--noise rate, with a $51~\mu\mbox{s}$ deadtime to 
suppress afterpulses, on String 21 is less
than $400~\mbox{Hz}$ per DOM, with somewhat higher rates in the deepest modules.
The time calibration system has been verified and shown to have an intrinsic 
resolution of less than $3~\mbox{ns}$. The overall timing resolution has been
studied with light sources on the DOM and seen to be as good as $3~\mbox{ns}$.

Muon tracks have also been reconstructed. The hit multiplicity and zenith angle
distributions match expectations from simulated atmospheric muons. 
The space--angle resolution, with one string, is found to
vary from $9.7^\circ$, for low multiplicity events, to $1.5^\circ$, for high
multiplicity events.

Finally, a search for upward--going events resulted in two candidate events in
six months of data. The events had multiplicities of 35 and 50 (out of 60) and 
zenith angles of $178.8^\circ$ and $179.1^\circ$, respectively
(errors are statistical only). Two events in
this time period is consistent with predictions from simulation~\cite{sim}.

The verification of the 9--string, 16--station array is now underway. More than
99\% of the modules are operational and preliminary analysis indicates that the
detector is functioning as designed.

\section{Conclusion}
The AMANDA detector continues the exploration of the high energy neutrino sky
and will soon be an integral part of its successor, IceCube. IceCube has
deployed 9 in--ice strings and 16 surface stations and early results are
promising.

\section*{Acknowledgments}
In addition to acknowledging the sources listed at the URL given in
footnote~(\ref{foot}), the author wishes to acknowledge NSF award number 
PHY-0611671 and EU contract number MSCF-CT-2004-516636 for support in attending
this conference.

\end{document}